\documentclass{article}

\renewcommand{\arraystretch}{1.5} 
\usepackage{amsmath,amssymb,booktabs}
\usepackage[margin=1in, right=1.5in, verbose=true,letterpaper]{geometry}  
\usepackage{multirow}
\usepackage{longtable}
\usepackage{tabularx}
\usepackage{PRIMEarxiv}
\usepackage{amssymb}
\usepackage[utf8]{inputenc} 
\usepackage[T1]{fontenc}    
\usepackage{hyperref}       
\usepackage{url}            
\usepackage{booktabs}       
\usepackage{amsfonts}       
\usepackage{nicefrac}       
\usepackage{microtype}      
\usepackage{lipsum}
\usepackage{fancyhdr}       
\usepackage{graphicx}       
\graphicspath{{media/}}     
\usepackage{hyperref} 
\usepackage[table]{xcolor}
\usepackage{colortbl}
\usepackage{tikz}
\usetikzlibrary{shapes, arrows, positioning, fit}
\usepackage{graphicx}
\usepackage{subfigure} 
\usepackage{threeparttable}
\title{When Pigs Get Sick: Multi-Agent AI \\ for Swine Disease Detection}

\author{
  Tittaya Mairittha,
  Tanakon Sawanglok,
  Panuwit Raden, 
  Sorrawit Treesuk\\
  AXONS \\
  \texttt{aimlstaff@axonstech.com} \\
}

\begin{document}
\maketitle

\begin{abstract}
Swine disease surveillance is critical to the sustainability of global agriculture, yet its effectiveness is frequently undermined by limited veterinary resources, delayed identification of cases, and variability in diagnostic accuracy. To overcome these barriers, we introduce a novel AI-powered, multi-agent diagnostic system that leverages Retrieval-Augmented Generation (RAG) to deliver timely, evidence-based disease detection and clinical guidance. By automatically classifying user inputs into either Knowledge Retrieval Queries or Symptom-Based Diagnostic Queries, the system ensures targeted information retrieval and facilitates precise diagnostic reasoning. An adaptive questioning protocol systematically collects relevant clinical signs, while a confidence-weighted decision fusion mechanism integrates multiple diagnostic hypotheses to generate robust disease predictions and treatment recommendations. Comprehensive evaluations encompassing query classification, disease diagnosis, and knowledge retrieval demonstrate that the system achieves high accuracy, rapid response times, and consistent reliability. By providing a scalable, AI-driven diagnostic framework, this approach enhances veterinary decision-making, advances sustainable livestock management practices, and contributes substantively to the realization of global food security. 
\end{abstract}

\section{Introduction}\label{sec:introduction}

Efficient surveillance and management of swine diseases are vital for maintaining global agricultural productivity, ensuring food security, and safeguarding animal welfare. However, the shortage of specialized veterinary expertise, especially in resource-limited regions, often results in delayed disease detection and ineffective responses. These challenges exacerbate economic losses, disrupt food supply chains, and negatively impact overall animal health~\cite{vanderwaal2018global}. Consequently, the timely and accurate identification of diseases is fundamental for the stability and sustainability of the swine industry.

To address this critical expertise gap, we introduce an AI-powered multi-agent diagnostic platform that utilizes Retrieval-Augmented Generation (RAG). The system supports veterinarians, animal health professionals, and swine husbandry professionals by delivering precise and context-sensitive diagnostic insights. The diagnostic process begins with an advanced query-classification methodology, categorizing submitted queries into two primary types: Knowledge Retrieval Queries, addressing general swine health information such as preventive care, vaccination strategies, medication guidelines, and biosecurity best practices, and Symptom-Based Diagnostic Queries, requiring comprehensive analysis of symptom descriptions to generate accurate differential diagnoses and targeted treatment recommendations.

A core innovation of our platform is its adaptive questioning mechanism, which dynamically collects detailed symptom data to refine diagnostic accuracy. Disease predictions are further enhanced through a confidence-weighted decision fusion approach, which prioritizes high-certainty results while flagging cases that require further expert evaluation. Additionally, the system provides actionable guidance, including disease control recommendations, sample collection protocols, treatment instructions, and preventive measures, ensuring its practical applicability in swine health management.

The effectiveness of our diagnostic approach was validated through a rigorous evaluation framework integrating self-correcting Large Language Models (LLMs) to minimize hallucinations and utilizing a curated domain-specific corpus to improve diagnostic precision and contextual reliability. Performance benchmarking was conducted across four critical swine diseases: African Swine Fever (ASF), Porcine Reproductive and Respiratory Syndrome (PRRS), Porcine Epidemic Diarrhea (PED), and Foot-and-Mouth Disease (FMD). Continuous expert-driven feedback and iterative refinements significantly enhanced the system’s diagnostic accuracy, clinical relevance, and overall practical utility.

By bridging the veterinary expertise gap with scalable AI-driven diagnostics, our platform empowers animal health professionals to make informed, timely decisions. This advancement strengthens disease surveillance, enhances animal welfare, and bolsters the resilience and sustainability of global pork production.

\section{Related Works}\label{sec:related_work}

Recent advancements in LLMs have significantly contributed to the development of multi-agent AI architectures, particularly for complex decision-making in healthcare applications. For instance, MDAgents~\cite{kim2024mdagents} dynamically adapt agent roles based on case complexity, while the AMSC framework~\cite{wang2024beyond} leverages probabilistic coordination among specialist agents to improve diagnostic accuracy. Additionally, researchers have explored deep reinforcement learning for personalized treatment optimization~\cite{shaik2023adaptive, lin2024multi}, ensemble learning for enhancing diagnostic precision~\cite{walia2024multi}, and knowledge-based AI frameworks that integrate structured clinical guidelines into decision-support systems~\cite{ke2024enhancing}. These developments highlight the superiority of multi-agent architectures over single-model AI systems, particularly in scalability, adaptability, and diagnostic consistency.

In veterinary medicine, AI-powered diagnostic and decision-support systems are emerging, albeit at an earlier stage than in human healthcare. Sobkowich~\cite{sobkowich2025demystifying} emphasizes the increasing role of AI in veterinary clinical workflows, including automated disease detection and medical imaging analysis, demonstrating how AI-driven tools streamline and enhance veterinary practice. Beyond diagnostics, machine learning models assist veterinarians in personalizing treatment strategies based on an animal’s genetic profile, medical history, and clinical symptoms~\cite{akinsulie2024potential}. Meanwhile, Joslyn et al.~\cite{joslyn2022evaluating} highlight the importance of collaborative AI models in veterinary radiology, arguing that multi-agent coordination—akin to human medical expert consultations—enhances diagnostic reliability in complex cases.

AI innovations in veterinary applications extend to specialized LLMs trained on clinical notes, which have been shown to outperform traditional disease prediction models~\cite{jiang2023vetllm}. Similarly, generative AI models have been applied to prescription analysis, automating medication dosage extraction and reducing prescription errors~\cite{hur2024is}. Additionally, multimodal AI frameworks such as MM-Vet integrate textual, imaging, and sensor data, expanding beyond text-based diagnostics to comprehensive AI-assisted decision-making~\cite{yu2024mmvet}. These advancements form a strong foundation for AI-driven decision support in veterinary medicine, with multi-agent methodologies poised to improve diagnostic accuracy, efficiency, and scalability.

In livestock farming, AI-powered precision monitoring systems are transforming disease surveillance and herd management. Even in traditional farm environments, multi-agent AI frameworks combined with IoT-based sensor networks have been implemented for real-time remote health monitoring~\cite{barriuso2018combination}. In swine disease management, AI-driven innovations are shaping next-generation multi-agent diagnostic systems. For example, deep learning models have been developed to interpret lateral flow assay results for ASF, providing rapid and sensitive field diagnostics for early outbreak detection~\cite{bakshi2025toward}.

AI-powered acoustic analysis has also been employed to automatically detect respiratory diseases in pigs based on cough patterns~\cite{lagua2023artificial}. Notably, AI cough recognition models can differentiate between productive, disease-related coughs and non-pathological environmental coughs caused by dust or irritants~\cite{benjamin2019precision}. Furthermore, AI-driven continuous cough monitoring, when combined with periodic pathogen testing, acts as an early-warning system for respiratory outbreaks. Field studies indicate that AI-monitored distress signals correlate with spikes in specific pathogens, enabling proactive disease containment before traditional detection methods confirm an outbreak~\cite{eddicks2024monitoring}.

While cough recognition AI focuses on respiratory disease detection, computer vision techniques provide an additional layer of disease surveillance by analyzing pig behavior and physiological changes. AI models analyzing video feeds have demonstrated the ability to detect subtle behavioral shifts—such as reduced movement, altered eating habits, or abnormal social interactions—often days before clinical symptoms appear~\cite{ryu2022object, wang2022research}. These automated surveillance tools address key bottlenecks in manual disease monitoring, which is labor-intensive and susceptible to observational errors. By automating clinical sign detection and aggregating herd data, AI-based systems significantly improve the speed, accuracy, and scalability of swine disease surveillance.

This study refines multi-agent AI frameworks to address swine disease detection challenges, integrating proven methodologies from human healthcare with veterinary diagnostics. By coordinating specialized agents—such as domain knowledge retrievers and diagnostic decision-makers—the system enables timely, evidence-based veterinary support. It enhances early disease identification, guides targeted interventions, and mitigates economic losses. Ultimately, this research contributes to a more resilient livestock sector, where AI-driven surveillance strengthens veterinary expertise and supports global food security.

\section{Methodology}

The system follows a structured three-stage methodology, beginning with \textbf{Query Classification~\ref{subsec:query_cls}}, which categorizes user queries to ensure appropriate processing. Next, \textbf{Multi-Agent Disease Prediction~\ref{subsec:multi_agent_predict}} utilizes specialized agents to analyze symptom-based queries, applying confidence-weighted decision fusion to identify the most probable diagnoses. Finally, \textbf{Recommendations~\ref{subsec:recommendations}} leverage RAG to provide targeted veterinary guidance.

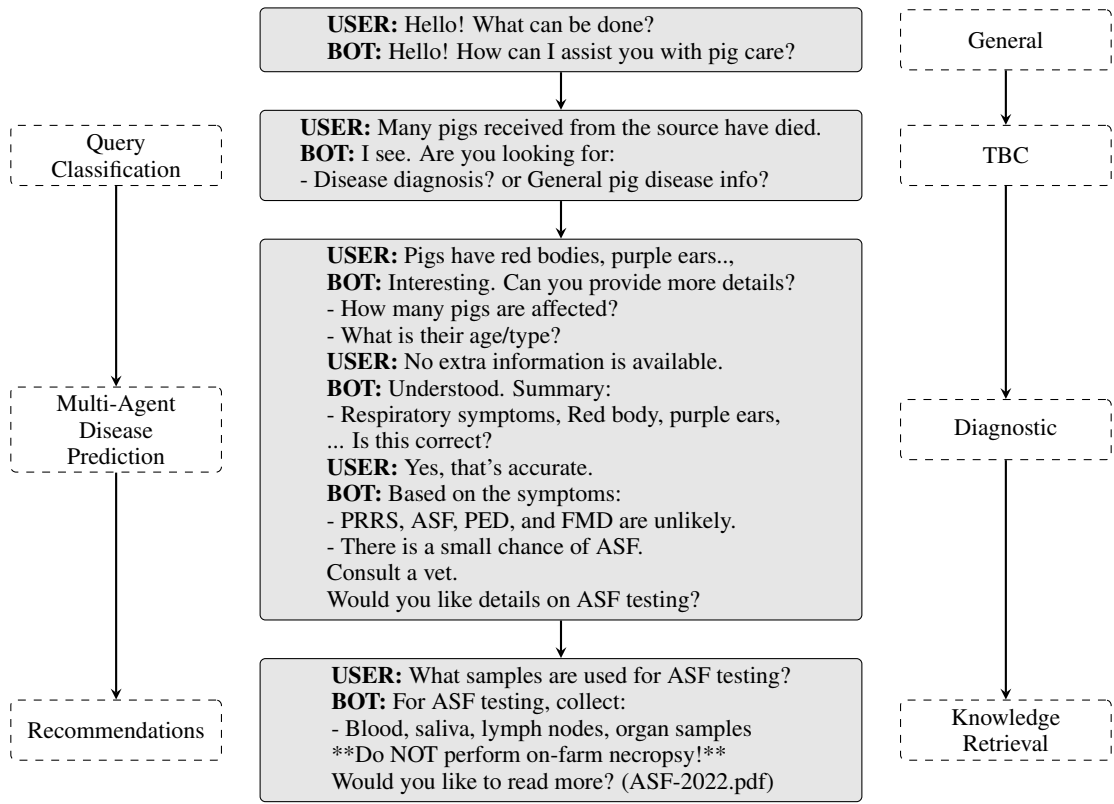
\begin{figure}[htbp]
    \centering
    \footnotesize
    \vspace{0.5cm}
    \begin{tikzpicture}[
        node distance=0.5cm,
        every node/.style={align=left},
        process/.style={minimum width=2.8cm, minimum height=0.8cm, align=center, rounded corners=2pt, draw, dashed},
        arrow/.style={thick, ->, >=stealth},
        dashedbox/.style={draw, rounded corners=2pt, minimum width=8cm, fill=gray!20}
    ]

    \node[dashedbox] (G) {
        \textbf{USER:} Hello! What can be done? \\
        \textbf{BOT:} Hello! How can I assist you with pig care?
    };
    
    \node[dashedbox, below=of G] (T) {
        \textbf{USER:} Many pigs received from the source have died. \\
        \textbf{BOT:} I see. Are you looking for: \\
        - Disease diagnosis? or General pig disease info?
    };
    
    \node[dashedbox, below=of T] (D) {
        \textbf{USER:} Pigs have red bodies, purple ears.., \\
        \textbf{BOT:} Interesting. Can you provide more details? \\
        - How many pigs are affected? \\
        - What is their age/type? \\
        \textbf{USER:} No extra information is available. \\
        \textbf{BOT:} Understood. Summary: \\
        - Respiratory symptoms, Red body, purple ears, \\ 
        ... Is this correct? \\
        \textbf{USER:} Yes, that's accurate. \\
        \textbf{BOT:} Based on the symptoms: \\
        - PRRS, ASF, PED, and FMD are unlikely. \\
        - There is a small chance of ASF. \\
        Consult a vet. \\
        Would you like details on ASF testing?
    };
    
    \node[dashedbox, below=of D] (K) {
        \textbf{USER:} What samples are used for ASF testing? \\
        \textbf{BOT:} For ASF testing, collect: \\
        - Blood, saliva, lymph nodes, organ samples \\
        **Do NOT perform on-farm necropsy!** \\
        Would you like to read more? (ASF-2022.pdf)
    };

    \draw[arrow] (G.south) -- (T.north);
    \draw[arrow] (T.south) -- (D.north);
    \draw[arrow] (D.south) -- (K.north);

    \node[process, left=of T] (QR) {Query \\ Classification};
    \node[process, left=of D] (MP) {Multi-Agent \\ Disease \\ Prediction};
    \node[process, left=of K] (ADM) {Recommendations};

    \draw[arrow] (QR.south) -- (MP.north);
    \draw[arrow] (MP.south) -- (ADM.north);

    \node[process, right=of G] (GS) {General};
    \node[process, right=of T] (TBCS) {TBC};
    \node[process, right=of D] (DS) {Diagnostic};
    \node[process, right=of K] (AS) {Knowledge \\ Retrieval};

    \draw[arrow] (GS.south) -- (TBCS.north);
    \draw[arrow] (TBCS.south) -- (DS.north);
    \draw[arrow] (DS.south) -- (AS.north);

    \end{tikzpicture}

   \vspace{0.3cm}
    \caption{Schematic representation of the multi-turn diagnostic conversation flow. This diagram illustrates how user queries transition from initial query classification to subsequent processing by specialized modules— including multi-agent disease prediction and RAG—ultimately leading to actionable veterinary recommendations.}
    \label{fig:stage_wise_conversation}
\end{figure}

\subsection{Query Classification}\label{subsec:query_cls}

To enhance diagnostic accuracy and response efficiency, user queries are first categorized into four classes:
\[
   \mathcal{C} \;=\; \{\text{K},\; \text{D},\; \text{T},\; \text{G}\}.
\]
Given a user query \(q\), a classification model provides probabilities \(P(c \mid q)\) for each \(c \in \mathcal{C}\). 
The system then chooses
\[
   \hat{c} \;=\; \arg\max_{c \,\in\, \mathcal{C}} \; P(c \mid q),
\]
and routes the query accordingly:

\begin{itemize}
 \item \textbf{K (Knowledge Retrieval Queries)}:\label{subsubsection:kb_query} These queries seek factual, evidence-based information on swine health topics, including disease prevention, vaccination protocols, prescriptions, and biosecurity measures. They correspond to the Recommendations stage~\ref{subsec:recommendations}, meaning users can directly receive guidance without first undergoing the symptom-based diagnostic process. Instead, K queries are processed directly through RAG, enabling the system to provide immediate and actionable recommendations.

\item \textbf{D (Symptom-Based Diagnostic Queries)}: \label{subsubsection:symtom_query}
These queries describe observed clinical signs in pigs and are analyzed by the system's diagnostic model to generate differential diagnoses and treatment recommendations. The system collects symptom data in multiple steps before predicting the most likely diseases.

\item \textbf{T (TBC Queries)}: 
TBC (to-be-clarified) queries are ambiguous, vague, or incomplete inputs. While they do not directly contribute to disease diagnostics, they prompt the user for additional details enabling accurate query classification. Effectively handling TBC queries enhances system adaptability and guides users toward more structured, actionable inputs.

\item \textbf{G (General Queries)}: 
General queries include casual greetings, off-topic questions, or broad farm-related inquiries not directly related to disease. While they do not contribute to disease diagnostics, they help maintain a natural conversational flow and ensure the system can manage incomplete or irrelevant inputs.
\end{itemize}

\subsection{Multi-Agent Disease Prediction}\label{subsec:multi_agent_predict}

Once a query is identified as a Symptom-Based Diagnostic, the system follows a structured process for disease prediction.

\subsubsection{Stage-Wise Symptom Collection}
The system defines three key states---\textbf{General} 
(\(G\)), \textbf{External} (\(E\)), and \textbf{Specific}~(\(S\))—and limits the symptom collection process to a maximum of three user–system exchanges before proceeding to disease prediction:

\[
  \mathcal{S} \;=\; \{\,G,\,E,\,S\}.
\]
\begin{itemize}
  \item \textbf{State \(\boldsymbol{G}\) (General):} 
  Broad health indicators such as mortality/morbidity rates, pig classification (piglets, breeders, 
  finishers), and environmental factors (ventilation, weather changes).

  \item \textbf{State \(\boldsymbol{E}\) (External):}
  Visible physical signs: skin lesions, color changes, nasal or ocular discharge, and notable 
  behavioral changes (e.g., refusal to stand, aggression).

  \item \textbf{State \(\boldsymbol{S}\) (Specific):}
  Targeted symptom clusters like respiratory (coughing, sneezing), gastrointestinal (vomiting, 
  diarrhea), neurological (tremors, seizures), or reproductive (stillbirths, infertility).
\end{itemize}

\subsubsection{State Transitions}
At each turn \(i\), the system maintains a state \(s_i \in S\), where the user's response \(a_i\) determines the next state via the transition function:
\[
   s_{i+1} \;=\; T\bigl(s_i, a_i\bigr),
\]
where \(T\) allows:
\begin{itemize}
  \item Sequential progression: \(G \to E \to S\).
  \item Direct transition: \(G \to S\) if highly specific details are provided early. 
  \item Reversion: \(S \to G\) if crucial data is missing and turns remain.
\end{itemize}
If the user provides no further input or the turn limit is reached, the system finalizes the collected data and initiates disease prediction (Figure~\ref{fig:stage_wise_conversation}).

\subsubsection{Confidence-Based Disease Prediction}

Each specialized disease agent \(i\) then assigns a confidence score \(p_i(D)\) for disease \(D\), aggregated as:

\[
   C(D) \;=\; \sum_{i=1}^{n} \alpha_i\, p_i(D),
\]

where \(\alpha_i\) represents the weight assigned to agent \(i\), typically reflecting its expertise or reliability. A disease is predicted when its confidence score meets or exceeds the predefined threshold \(\tau\):

\[
   \widehat{D} \;=\; \{\,D \mid C(D) \,\ge\, \tau\}.
\]

The system categorizes predictions into four confidence tiers to determine the appropriate response:
\begin{itemize}
  \item \textbf{Very High Confidence} (\(C(D) \ge 0.75\)): Strong predictive certainty; prioritized for action.
  \item \textbf{High Confidence} (\(0.624 \le C(D) < 0.75\)): Highly reliable but may require additional validation.
  \item \textbf{Medium Confidence} (\(0.375 \le C(D) < 0.624\)): Moderate likelihood; further assessment is advised.
  \item \textbf{Low Confidence} (\(C(D) < 0.375\)): Uncertain prediction; flagged for additional verification.
\end{itemize}

If no disease surpasses the confidence threshold \(\tau\) (set dynamically, e.g., 0.375 or 0.75 depending on usage context), the case is flagged as Out-of-Distribution (OOD), triggering an escalation protocol. In such cases, the system either: (1) alerts a veterinary expert for manual review, or (2) recommends additional diagnostic tests (e.g., blood samples, pathogen screening) before reaching a final diagnosis.

\subsection{Recommendations}\label{subsec:recommendations}

The Recommendations module employs RAG techniques integrated with LLM-based reasoning to deliver context-aware, evidence-based veterinary guidance. This approach ensures that diagnostic outcomes are transformed into clear, actionable recommendations, providing users with precise and scientifically grounded advice.

\subsubsection{Algorithmic Framework}
To ensure recommendations are context-aware and evidence-driven, the system employs a probabilistic approach, formulated as follows:
\[
    P(R \mid D) = \sum_{z \in \mathcal{Z}} P(R \mid D, z) \, P(z \mid D),
\]
where:
\begin{itemize}
    \item \(D\) denotes the diagnosed disease,
    \item \(R\) represents the generated recommendation,
    \item \(z\) is a document retrieved from the veterinary knowledge base \(\mathcal{Z}\),
    \item \(P(z \mid D)\) represents the probability of retrieving document \(z\) given \(D\),
    \item \(P(R \mid D, z)\) denotes the probability of generating \(R\) conditioned on both \(D\) and \(z\).
\end{itemize}

To ensure robust performance, an exponential backoff strategy with five steps is integrated into the LLM-based generation process. If an initial LLM call does not yield a satisfactory result, the system automatically retries the request with progressively longer delay intervals. This iterative process continues until \(P(R \mid D, z)\) is sufficiently refined to produce a reliable and actionable recommendation.

\subsubsection{Integrated Recommendation Pipeline}
Building on the algorithmic foundation described above, the pipeline is structured into several key components, each designed to translate theoretical principles into practical steps:

\begin{itemize}
    \item \textbf{Registration State:}
    Upon receiving a diagnosis \(D\) along with the corresponding user query, the system establishes a task-specific state that encapsulates query details, conversation history, and relevant metadata. Concurrently, a tailored filter expression is formulated to guide the document retrieval process.

    \item \textbf{Entity Extraction:}
    Two complementary retrieval mechanisms are employed:
    \begin{itemize}
    \item General Entity Extraction: Identifies key veterinary terms and domain-specific terminology used by experts. This includes specialized disease references, such as ``Roll Over,'' which describes a swine health status transition from disease-free to infected. These entities are retrieved from the knowledge base relevant to \(D\) to ensure accurate context.
    \item Medicine and Vaccine-Specific Extraction: Focuses on isolating pharmaceutical and vaccine-related information, such as trade names and medicine or vaccine groups, ensuring precise medical guidance in recommendations.
    \end{itemize}

    \item \textbf{Query Contextualization:}
    A two-stage rewriting process refines the original query. First, general entities and conversation history are used to construct a detailed query. Then, vaccine-specific information is integrated, enhancing query precision.

    \item \textbf{Document Retrieval and Processing:}
    The refined query is used to retrieve documents from a vectorized data store. Extracted documents, along with their metadata, are consolidated to establish a robust context for the final reasoning phase.

    \item \textbf{LLM-Based Answer Generation:}
    The comprehensive context—augmented with historical interactions and other relevant parameters—is provided as input to an LLM. An exponential backoff strategy ensures resilient and reliable LLM call management.

    \item \textbf{Output Assembly:}
    Finally, the system aggregates the generated recommendation along with retrieved documents and associated metadata. The final output is presented to the user in a clear and actionable format.
\end{itemize}

\section{Experimental Setup}\label{sec:experimental_setup}
The AI system's performance is evaluated across three tasks: Query Classification, Disease Diagnosis, and Knowledge Retrieval. These tasks assess its ability to classify queries, diagnose diseases accurately, and retrieve relevant medical knowledge. The evaluation framework ensures precision and reliability by measuring accuracy, effectiveness, and relevance. The experimental design separates training, validation, and test datasets to evaluate both internal consistency and external generalizability.

\subsection{Data Collection}
To maintain consistency and facilitate precise comparisons across tasks, dataset details are summarized in a single table (Table~\ref{tab:dataset_summary}). This table provides a structured summary of the distribution of validation and test questions for each task, ensuring clarity and uniformity in data representation.

\begin{table}[h]
\centering
\small
\begin{tabular}{|l|c|c|}
\hline
\textbf{Task} & \textbf{Validation Questions} & \textbf{Test Questions} \\
\hline
Query Classification & 220 & 461 \\
Disease Diagnosis & 52 & 32 \\
Disease-Related Knowledge Retrieval & 2,808 & 128 \\
Vaccine-Related Knowledge Retrieval & 906 & 191 \\
\hline
\rowcolor{gray!20} \textbf{Total} & \textbf{3,986} & \textbf{812} \\
\hline
\end{tabular}
\vspace{0.3cm}
\caption{Comprehensive dataset summary detailing the total number of validation and test questions across tasks.}
\label{tab:dataset_summary}
\end{table}

Both the validation and test sets are constructed using the same methodology, with the primary distinction being their source of origin:  
\begin{itemize} 
\item \textbf{Validation Set:} AI-generated following structured guidelines.   
\item \textbf{Test Set:} Curated by veterinary experts.  
\end{itemize}

The questions were systematically derived from relevant references, including:  
\begin{itemize} 
\item \textbf{Disease-related references} (e.g., ASF, PRRS, FMD, PED).  
\item \textbf{Vaccine-related references} (e.g., vitamins, disinfectants, hormones, antihelminthics).  
\end{itemize}

Each question was carefully annotated with the following metadata:  
\begin{itemize} 
\item \textbf{Scenario:} The specific scenario or methodology used to generate the question (e.g., multi-turn dialogue, n-page document analysis).  
\item \textbf{Question Type:} Categorized by complexity and structure (e.g., factoid, yes–no).  
\item \textbf{Document Source:} The exact document and page number used as the basis for question generation.  
\item \textbf{Example Answer:} A model response demonstrating the expected level of detail and accuracy.  
\end{itemize}

This structured approach to data collection ensures that both datasets adhere to a uniform format and difficulty distribution, enabling a fair and rigorous evaluation of model performance across different question types and complexities.

\subsection{Evaluation Methodology}\label{sec:evaluation_methodology}

This section outlines the evaluation framework for assessing our AI system's performance in Query Classification, Disease Diagnosis, and Knowledge Retrieval. We define key metrics for each task and present a comparative performance analysis.

\subsubsection{Evaluation Metrics for Query Classification and Routing}

For the query classification task, we evaluate the system using the following metrics:
\begin{itemize}
    \item \textbf{Precision}: The proportion of correctly predicted positive cases out of all cases predicted as positive.
    \item \textbf{Recall}: The proportion of correctly predicted positive cases out of all actual positive cases.
    \item \textbf{F1-score}: The harmonic mean of precision and recall.
    \item \textbf{Accuracy}: The overall proportion of correctly classified cases within the dataset.
\end{itemize}

\subsubsection{Evaluation Metrics for Symptom-Based Diagnostic Queries}

The disease diagnosis task is evaluated using a set of metrics that capture both the accuracy of the diagnostic predictions and the efficiency of the response generation:
\begin{itemize}
    \item \textbf{Accuracy}: Models were evaluated based on diagnostic accuracy, defined as the percentage of cases in which the correct diagnosis appeared among the top two ranked predictions.
    \item \textbf{Execution Time}: The time (in seconds) taken by the model to produce its diagnostic output, reflecting computational efficiency.
\end{itemize}

\textbf{Comparative Analysis:}
In our pilot study, we selected three leading models: o1-mini, GPT-4o, and Gemini-1.5-Pro-002—for disease diagnosis. Their selection is supported by:
\begin{itemize}
    \item \textbf{o1-mini:} Exhibits efficient language understanding and rapid context retention, making it suitable for streamlined diagnostic applications~\cite{temsah2024openai}.
    \item \textbf{GPT-4o:} Provides advanced integration of diverse clinical data, with its capabilities outlined in recent evaluations~\cite{zhang2024latest}.
    \item \textbf{Gemini-1.5-Pro-002:} Employs a multimodal approach tailored for medical-specific inputs, showing promising diagnostic precision in early assessments~\cite{saab2024capabilities}.
\end{itemize}

\subsubsection{Evaluation Metrics for Knowledge Retrieval Queries}

For the knowledge retrieval task, responses are evaluated using a tailored scoring system (ranging from 0 to 5) based on the following dimensions:
\begin{itemize}
    \item \textbf{Expansiveness}: The extent to which the response thoroughly covers the topic.
    \item \textbf{Coherence}: The logical organization and readability of the response.
    \item \textbf{Correctness}: The assessment of whether the response is overall accurate and well-supported by evidence.
    \item \textbf{Relevance}: The degree to which the response directly addresses the query.
    \item \textbf{Accuracy}: The correctness of the response based on the question type, ensuring it is free from misinformation.
\end{itemize}

\textbf{Comparative Analysis:}
To evaluate improvements in our knowledge retrieval system, we compared it against the approach in~\cite{vertex_ai_search}, a global platform for generalized retrieval tasks. Using validation and test datasets, we analyzed average performance scores and conducted paired t-tests on matched samples with a 0.05 significance threshold. To enhance reliability, we applied bootstrapping, sampling 80\% of matched data points repeatedly and averaging the t-statistics and p-values.

\section{Results}\label{sec:results}
\subsection{Query Classification Results}\label{sec:result_cls}

The table~\ref{tab:confusion_matrix} presents the classification performance of the system across four query types: General, Retrieval (Knowledge Retrieval), Diagnosis (Symptom-Based Diagnostic), and TBC. The system achieves an overall accuracy of \textbf{95.23\%}, demonstrating strong performance in correctly categorizing user queries.

\begin{table}[ht]
\centering
\small
\renewcommand{\arraystretch}{1.3}
\begin{tabular}{|c|c|c|c|c|c|c|c|}
\hline
\multirow{2}{*}{\textbf{Ground Truth}} & \multicolumn{4}{c|}{\textbf{Prediction}} & \multirow{2}{*}{\textbf{Precision}} & \multirow{2}{*}{\textbf{Recall}} & \multirow{2}{*}{\textbf{F1-score}} \\
\cline{2-5}
 & \textbf{General} & \textbf{Retrieval} & \textbf{Diagnosis} & \textbf{TBC} &  &  &  \\
\hline
\textbf{General}   & \textbf{24}  & 0  & 0  & 0  & 0.750  & \textbf{1.000}$^\dagger$  & 0.857  \\ \hline
\textbf{Retrieval}    & 6  & \textbf{301} & 5 & 6  & \textbf{0.993}$^\dagger$  & 0.947  & \textbf{0.969}$^\dagger$  \\ \hline
\textbf{Diagnosis} & 0  & 2  & \textbf{82} & 1  & 0.943  & 0.965  & 0.953  \\ \hline
\textbf{TBC}    & 2  & 0  & 0  & \textbf{32}  & 0.821  & 0.941  & 0.877  \\ \hline
\multicolumn{5}{|c|}{} & \multicolumn{3}{c|}{\cellcolor{black!10} \textbf{Accuracy : 95.23\%}} \\ \hline
\end{tabular}
\vspace{0.3cm}
\caption{Classification performance on the test set. The best values are marked with $^\dagger$.}
\label{tab:confusion_matrix}
\end{table}

The classification system performs well across different query types. General queries maintain a perfect recall of 1.000, ensuring all such queries are correctly identified, though the precision is slightly lower at 0.750 due to some misclassifications. This results in an F1-score of 0.857, indicating a good balance between precision and recall, although some ambiguous queries were mistakenly classified as General instead of TBC.  

Retrieval queries exhibit exceptional performance, with a high precision of 0.993 and recall of 0.947, meaning the system efficiently retrieves factual information. However, a few cases were misclassified, particularly as symptom-based diagnostic queries.  

Diagnosis queries achieve a precision of 0.943 and a recall of 0.965, signifying that most symptom-based queries are correctly identified. Minor misclassifications occur, particularly into Retrieval queries, when symptom descriptions resemble knowledge-seeking questions. Despite this, the F1-score of 0.953 reflects strong diagnostic capability.  

Lastly, TBC queries are well-handled, with a precision of 0.821 and a recall of 0.941, demonstrating the system’s effectiveness in managing ambiguous and incomplete queries. However, some TBC queries may be mistakenly classified as General or Diagnosis due to insufficient context in user input.

\subsection{Symptom-Based Diagnostic Results}
\label{sec:diagnosis_results}

As shown in Table~\ref{tab:diagnosis_combined}, although all models achieve similar test accuracy, execution times differ significantly. Notably, \textbf{GPT-4o} achieves the highest test accuracy at \textbf{90.63\%} while also being the fastest, with an execution time of \textbf{18.78} seconds. Its validation accuracy of 88.46\% remains close, demonstrating consistency and robustness across datasets. In contrast, Gemini-1.5-Pro-002 achieves the highest validation accuracy at 94.23\% but drops to 87.50\% on the test set, suggesting potential generalization issues. Meanwhile, o1-mini-mini lags in both accuracy and speed, with the highest latency at 29.38 seconds, which limits its practical usability. Additionally, ASF and PED yield the most reliable diagnoses, whereas PRRS and FMD exhibit inconsistencies. A deeper analysis in Section~\ref{sec:incorrect_prediction} will explore potential reasons for these variations.

\begin{table}[ht]
\centering
\small
\renewcommand{\arraystretch}{1.3}
\begin{tabular}{|l|c|c|c|c|c|c|}
\hline
\multirow{2}{*}{\textbf{Model}} & \multirow{2}{*}{\textbf{Disease}} & \multicolumn{2}{c|}{\textbf{Predicted / Actual}} & \multicolumn{2}{c|}{\textbf{Execution Time (s)}} \\
\cline{3-6}
 &  & \textbf{Validation} & \textbf{Test} & \textbf{Validation} & \textbf{Test} \\
\hline
\multirow{5}{*}{\textbf{GPT-4o}} 
 & ASF  & (7/7)  & (5/5)   & 20.09  & 17.83  \\
 & PRRS & (4/5)  & (11/14)    & 19.20  & 18.82  \\
 & PED  & (4/4)  & (5/5)   & 19.02  & 20.17  \\
 & FMD  & (3/6)   & (7/7)    & 19.94  & 18.75  \\
 & OOD  & (28/30)   & (1/1)   & 19.53  & 18.31  \\
\hline
\rowcolor{gray!20} \textbf{Accuracy (\%)} &  & 88.46 & \textbf{90.63}$^\dagger$  & \textbf{19.55}$^\dagger$ & \textbf{18.78}$^\dagger$  \\
\hline
\multirow{5}{*}{\textbf{o1-mini-mini}} 
 & ASF  & (7/7)  & (5/5)  & 27.69  & 28.55  \\
 & PRRS & (5/5)  & (10/14)   & 31.03  & 29.19  \\
 & PED  & (4/4)  & (5/5)  & 31.38  & 31.24  \\
 & FMD  & (3/6)  & (6/7)  & 27.89  & 27.57  \\
 & OOD  & (24/30)   & (1/1)  & 28.91  & 26.57  \\
\hline
\rowcolor{gray!20} \textbf{Accuracy (\%)} & & 82.69 & 84.37 & 29.38 & 28.63 \\
\hline
\multirow{5}{*}{\textbf{Gemini-1.5}} 
 & ASF  & (7/7)  & (5/5)  & 22.19  & 21.87  \\
 & PRRS & (5/5)  & (11/14)   & 22.42  & 24.62  \\
 & PED  & (4/4)  & (5/5)  & 21.58  & 26.31  \\
 & FMD  & (3/6)   & (6/7)   & 23.52  & 20.62  \\
 & OOD  & (30/30)  & (1/1)  & 21.82  & 20.30  \\
\hline
\rowcolor{gray!20} \textbf{Accuracy (\%)} &  & \textbf{94.23}$^\dagger$ & 87.50 & 22.30 & 22.74 \\
\hline
\end{tabular}
\vspace{0.3cm}
\caption{Performance comparison of different models on disease diagnosis tasks.}
\label{tab:diagnosis_combined}
\end{table}

\begin{figure}[ht]
    \centering
    \subfigure[Confidence Distribution for Correct Predictions]{
        \includegraphics[width=0.45\linewidth]{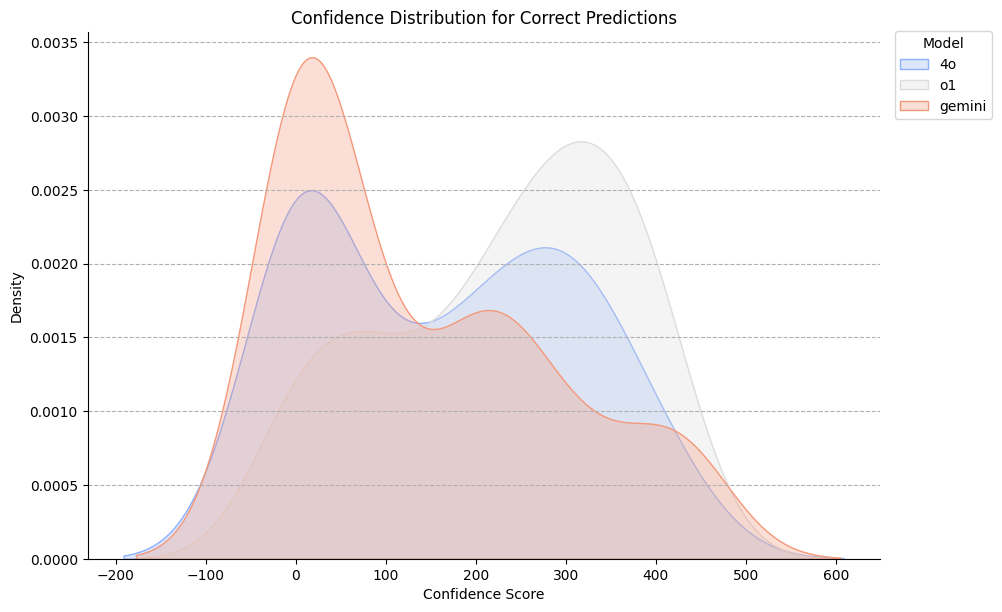}
        \label{fig:cof_distribution}
    }
    \subfigure[Score Distributions across Key Performance Metrics]{
        \includegraphics[width=0.45\linewidth]{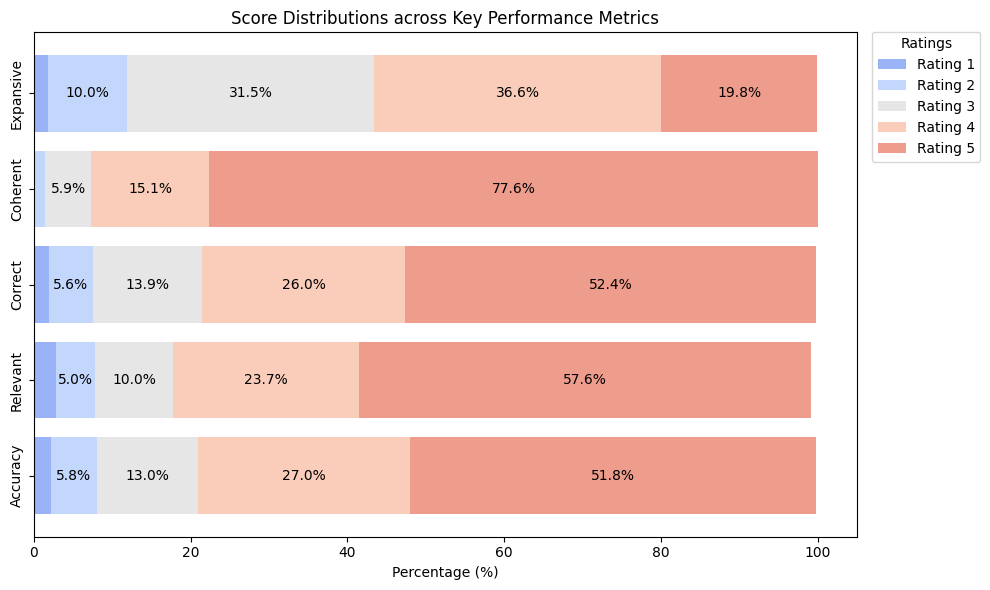}
        \label{fig:score_distribution}
    }
    \caption{Subfigure (a) illustrates the distribution of model confidence scores specifically for correct predictions, offering insights into the relative certainty of outputs across different models. Subfigure (b) presents the performance ratings distribution across key evaluation metrics, highlighting the model’s consistent and strong performance in accuracy, relevance, correctness, coherence, and expansiveness.}
    \label{fig:side-by-side}
\end{figure}

To further analyze model behavior, Figure~\ref{fig:cof_distribution} presents the confidence distribution of each model, highlighting differences in prediction stability. Models with higher peak densities exhibit more concentrated confidence levels, whereas broader distributions indicate greater variability.

Among the evaluated models, GPT-4o demonstrates the most balanced confidence distribution, striking an ideal middle ground between predictive consistency and adaptability. It maintains a moderate peak density with a controlled spread, ensuring stable and well-calibrated confidence scores across different predictions. o1-mini follows closely, exhibiting a slightly broader spread, which suggests higher variability but still maintains reasonable reliability.

In contrast, Gemini has the broadest confidence distribution despite its high peak density, indicating significant variability in confidence scores. This inconsistency may lead to overconfidence in certain cases, as its predictions fluctuate more widely than those of the other models, making its reliability less predictable.

Given these characteristics, GPT-4o emerges as the optimal choice, offering the best balance between predictive reliability and adaptability. Its ability to maintain stable confidence levels while minimizing extreme variations makes it well-suited for applications requiring both accuracy and robustness in real-time decision-making.

\subsection{Knowledge Retrieval Results}\label{sec:result_search}

Table~\ref{tab:comparison} presents a side-by-side evaluation of our approach against the baseline across multiple datasets and evaluation metrics. The results demonstrate that our method outperforms the baseline in all measured dimensions. The most notable improvement is in Coherence, which increases from 3.92 to 4.61, followed by Relevance, which rises from 3.19 to 4.15. Accuracy also shows substantial gains, particularly in the Disease domain, where it jumps from 1.89 to 4.14 in the validation set. Similarly, the overall Final Score improves significantly, increasing from \textbf{3.31} to \textbf{4.17}, highlighting the robustness and consistency of our enhanced method.

\begin{table}[ht]
    \centering
    \small
    \resizebox{\textwidth}{!}{
    \begin{tabular}{|l|l|cc|cc|cc|cc|cc|cc|}
        \hline
        \multirow{2}{*}{} & \multirow{2}{*}{\textbf{Dataset}} & \multicolumn{2}{c|}{\textbf{Accuracy}} & \multicolumn{2}{c|}{\textbf{Relevance}} & \multicolumn{2}{c|}{\textbf{Correctness}} & \multicolumn{2}{c|}{\textbf{Coherence}} & \multicolumn{2}{c|}{\textbf{Expansiveness}} & \multicolumn{2}{c|}{\textbf{Final Score}} \\
        \cline{3-14}
        & & Baseline & Ours & Baseline & Ours & Baseline & Ours & Baseline & Ours & Baseline & Ours & Baseline & Ours \\
        \hline
        \multirow{2}{*}{\textbf{Val}} & Vaccine & 4.12 & 4.46 & 4.22 & 4.49 & 4.15 & 4.47 & 4.51 & 4.74 & 4.03 & 4.17 & 4.20 & 4.46\\
        & Disease & 1.89 & 4.14 & 1.57 & 4.21 & 2.09 & 4.15 & 2.98 & 4.69 & 1.82 & 3.44 & 2.06 & 4.13 \\
        \hline
        \multirow{2}{*}{\textbf{Test}} & Vaccine & 3.50 & 4.04 & 3.62 & 4.04 & 3.47 & 4.04 & 4.10 & 4.49 & 3.27 & 3.80 & 3.57 & 4.08 \\
        & Disease & 3.32 & 3.84 & 3.36 & 3.85 & 3.42 & 3.83 & 4.07 & 4.52 & 2.76 & 3.42 & 3.39 & 4.02 \\
        \hline
        \rowcolor{gray!20}  \multicolumn{2}{|c|}{\textbf{Average}} & 3.21 & \textbf{4.12}$^\dagger$ & 3.19 & \textbf{4.15}$^\dagger$ & 3.28 & \textbf{4.12}$^\dagger$ & 3.92 &  \textbf{4.61}$^\dagger$ &  2.97 & \textbf{3.71}$^\dagger$ & 3.31 & \textbf{4.17}$^\dagger$ \\
        \hline
    \end{tabular}
    } 
    \vspace{0.3cm}
    \caption{Comparison of the baseline and our method across different datasets and multiple evaluation metrics.}
    \label{tab:comparison}
\end{table}

Table~\ref{tab:val_test_comparison} presents a statistical comparison of Ours vs. Baseline across five evaluation metrics for both Validation and Test datasets, using paired t-tests. The results indicate statistically significant differences across all metrics, with Accuracy consistently showing the strongest improvements across both datasets. Correctness and Relevance also demonstrate notable improvements. However, Coherence and Expansiveness show contrasting trends: while both exhibit strong improvements in Validation, their statistical significance is weaker in Test, suggesting that the observed differences may be more sensitive to dataset variability. The consistently high t-statistics confirm a meaningful performance difference between our approach and the baseline, though some metrics may generalize better than others.

\begin{table}[ht]
    \centering
    \small
    \renewcommand{\arraystretch}{1.2}
    \setlength{\tabcolsep}{10pt}
    \begin{tabular}{|l|cc|cc|}
        \hline
        \multirow{2}{*}{\textbf{Metric}} & \multicolumn{2}{c|}{\textbf{t-statistic}} & \multicolumn{2}{c|}{\textbf{p-value}} \\
        \cline{2-5}
        & Validation & Test & Validation & Test \\
        \hline
        Accuracy      & 4.6787 & 8.3334 & 0.0009 & 0.0010 \\
        Coherence      & 4.5289 & 6.9511 & 0.0003 & 0.0100 \\
        Correctness   & 3.5681 & 7.5433 & 0.0016 & 0.0044 \\
        Expansiveness & 4.7365 & 6.9230 & 0.0004 & 0.0103 \\
        Relevance      & 3.5721 & 9.2212 & 0.0014 & 0.0063 \\
        \hline
    \end{tabular}
    \vspace{0.3cm}
    \caption{Statistical comparison of Ours vs. Baseline on validation and test sets.}
    \label{tab:val_test_comparison}
\end{table}

To further illustrate how these improvements manifest across different rating scales, Figure~\ref{fig:score_distribution} presents rating distributions. The Coherence category shows a particularly high concentration of top-level ratings, with \textbf{77.6\%} of responses rated 5. In contrast, Expansiveness exhibits a broader distribution, indicating greater variability in response depth. While Correctness, Relevance, and Accuracy also demonstrate strong top-tier ratings, a noticeable proportion of responses still fall within the 3 and 4 range. 

Overall, these findings confirm that our refinements have significantly enhanced response accuracy, contextual relevance, and the model’s adaptability to real-world, domain-specific queries.
\section{Discussion}\label{sec:discussion}

The analysis identifies several critical areas contributing to errors in our AI diagnostic system, providing insights into opportunities for further improvement.

\subsection{Ambiguity in Query Classification}
Table~\ref{tab:confusion_matrix} highlights instances where queries are misclassified, often being incorrectly assigned to the general category instead of their intended classification. These errors often result from ambiguous abbreviations or vague phrasing. For example, queries such as ``Please provide usage instructions'' and ``Ratio'' lack sufficient context for precise categorization. Similarly, the query ``DLD for Agita'' contains the abbreviation DLD, which is intended to refer to the Department of Livestock Development but can be misinterpreted as Digital Learning Development. As a result, the system issues a default response: ``Sorry, I cannot provide information on this topic. However, I can assist with swine diseases, vaccine diagnostics, or treatments''. Enhancing context-aware query processing by integrating advanced Natural Language Understanding (NLU) techniques and a targeted Named Entity Recognition (NER) model capable of disambiguating domain-specific abbreviations, such as "DLD," as outlined in Section~\ref{subsec:recommendations}, could significantly improve query classification accuracy.

Additionally, five queries have been incorrectly classified under Diagnostic. For instance, the query ``Are the breeding pigs infected with African Swine Fever?'' includes disease-specific terminology, inadvertently triggering a diagnostic classification, even when the user’s intent is to retrieve general information. Similarly, six retrieval queries have been misclassified under TBC. One example is the query ``I encountered dark stool and stool with blood. What injection should I use?'', which prompts the system to seek clarification: ``Do you need information on diagnosing or treating dark and bloody stool in pigs? Please confirm''. While these misclassifications are not always critical, they introduce unnecessary friction in the user interaction.

\subsection{Misclassification of Diseases}\label{sec:incorrect_prediction}  
Table~\ref{tab:diagnosis_combined} reveals strong diagnostic accuracy for some diseases, such as ASF and PED, but greater variability for others, like FMD and PRRS. These inconsistencies may stem from uneven representation in the training data or the inherently subtle symptom profiles of certain diseases.  

In this case, the actual diagnosis was FMD, but the model sometimes predicted OOD. This could occur because the symptoms were too general and could be linked to other conditions, such as bacterial infections, nutritional deficiencies, or mechanical injuries, making it challenging for the model to precisely identify diseases. Additionally, low model confidence due to missing key diagnostic features may have led it to classify the case as OOD rather than making an uncertain disease assignment. In other instances, the model misclassified FMD as ASF, likely due to overlapping clinical signs such as lethargy and sudden death.

Another disease with a low ranking is PRRS. For example, in the case of: ``Abortions in sows, an increase in stillborn or mummified fetuses, respiratory symptoms, high fever, red skin, anorexia, and diarrhea'', the symptoms strongly align with PRRS; however, the system misclassified the case as OOD or ASF. The model likely considered the symptoms too broad or overlapping with multiple diseases, leading it to prioritize OOD due to uncertainty. Since ASF, PRRS, and FMD all cause fever, the system failed to recognize the reproductive and respiratory indicators as PRRS-specific. By over-prioritizing fever and redness, the model favored ASF while overlooking the strong reproductive symptoms of PRRS. As a result, PRRS was ranked lower than OOD and ASF, despite its clear symptom alignment.  

Many swine viral infections share overlapping symptoms, making diagnostic uncertainty a significant challenge. Symptoms such as fever, anorexia, diarrhea, and respiratory distress are common across multiple diseases, often resulting in misclassification. Additionally, because some diseases in the test set had very few cases (e.g., 1/1 or 5/5), the model’s performance for these diseases may not fully reflect its real-world generalization ability. This highlights the need for a more balanced dataset, particularly for underrepresented diseases, to provide a fairer and more robust evaluation of the model's diagnostic capabilities.  

\subsection{Low Confidence Predictions}\label{subsec:low_score} 
Despite overall improvements, some predictions received low scores (see Table~\ref{tab:comparison}), indicating areas where the system struggled with alignment, relevance, or completeness. Although fairness in evaluation is essential, a broader concern is the need for comprehensive responses fully capturing clinically relevant details.

\textbf{Example 1:} ASF Testing (Rating: 2)
\begin{itemize}
    \item \textbf{Query:} ``How does a lab test for ASF (African Swine Fever)?''
    \item \textbf{Prediction:} A general overview of ASF diagnostic methods, mentioning PCR and ELISA.
    \item \textbf{Ground Truth:} Detailed laboratory setup, including equipment and procedural steps for ASF testing.
    \item \textbf{Issue:} While the response correctly identifies PCR and ELISA as valid diagnostic methods, it lacks the detailed procedural and equipment-related information found in the reference. The lower score reflects incompleteness rather than inaccuracy, as the response omits but does not contradict essential details.
\end{itemize}

\textbf{Example 2:} Water Disinfection (Rating: 1)
\begin{itemize}
    \item \textbf{Query:} ``How to disinfect water before use?''
    \item \textbf{Prediction:} Describes the use of hydrogen peroxide for water disinfection.
    \item \textbf{Ground Truth:} Specifies chlorine-based disinfectants and correct dosage instructions.
    \item \textbf{Issue:} The Prediction presents a scientifically valid but alternative method not covered in the reference. The misalignment stems from the Ground Truth’s exclusive focus on chlorine-based disinfection, despite hydrogen peroxide being a legitimate option. The lower score reflects a lack of alignment rather than factual inaccuracy.
\end{itemize}

\textbf{Example 3:} Pig Diarrhea Treatment (Rating: 0.5)
\begin{itemize}
    \item \textbf{Query:} ``Which drug group can be used to treat black diarrhea in pigs?''
    \item \textbf{Prediction:} The system recommended Tiamulin as the treatment.
    \item \textbf{Ground Truth:} The reference document specifies Lincomycin and Spectinomycin as the correct treatment, along with dosage instructions.
    \item \textbf{Issue:} The Prediction suggests Tiamulin, a valid treatment not mentioned in the Ground Truth. While the reference lists Lincomycin and Spectinomycin, it does not explicitly exclude other treatments. The lower score reflects a mismatch with the reference rather than an actual error.
\end{itemize}

Analysis of the evaluation results indicates that certain factually correct yet alternative responses received lower scores primarily due to differences from the reference rather than genuine inaccuracies. This suggests that strict adherence to reference documents sometimes outweighs the recognition of clinically valid variations. While maintaining evaluation consistency remains important, responses should not only reflect alignment with reference materials but also emphasize completeness and clinical applicability. To address this, the knowledge retrieval component should incorporate a broader range of validated sources and expert-vetted alternatives, provide detailed procedural explanations, and recognize multiple clinically valid approaches. This strategy would enhance the depth and practical relevance of responses, ensuring scientific accuracy without overly restricting them to a single reference source.
\section{Conclusion and Future works}\label{sec:Conclusion}

This study demonstrates the robust capabilities of an AI-powered multi-agent diagnostic system designed to improve swine disease surveillance and veterinary decision-making. The system effectively distinguishes between different types of user queries—general, retrieval, diagnostic, and ambiguous—and shows notable strengths in diagnostic accuracy and response speed, particularly using the GPT-4o model. The knowledge retrieval module has also significantly advanced in accuracy, coherence, and practical applicability, surpassing previous methods.

However, certain areas remain open for further refinement. The overlap of clinical symptoms among swine diseases occasionally leads to diagnostic misclassification, suggesting that differential diagnostic capabilities can be further enhanced by improving symptom extraction precision and employing multimodal symptom analysis. Additionally, the system currently faces challenges when processing ambiguous or incomplete user queries. This underscores the need for more robust context-aware query interpretation mechanisms, better interactive clarification processes, and more sophisticated entity recognition models.

Improving response expansiveness and relevance is another critical direction for future research. Incorporating comprehensive explanations that include detailed procedural steps, multiple valid treatment options, and extensive clinical contexts will enhance the practical utility of generated responses. Establishing iterative expert-driven feedback loops will further ensure continuous improvement in accuracy and practical applicability.

To enhance real-time practical usability, optimizing the model inference efficiency through techniques such as compression, quantization, or streamlined architectures is also essential. Future work will additionally prioritize expanding the scope of disease coverage, particularly to address emerging and region-specific swine diseases. Integrating comprehensive veterinary procedural databases will further enhance the system’s practical utility by providing detailed, step-by-step guidance for veterinary procedures.

These refinements will further strengthen the system’s effectiveness, generalization capability, and practical usability, thereby significantly advancing veterinary medicine, animal health outcomes, and global agricultural sustainability.

\section{Acknowledgments}\label{sec:acknowledgments}

We sincerely thank the Swine Veterinary Service (SVS) team for identifying key pain points and providing valuable consultation, which helped shape the direction of this research. We are also grateful to SVS for curating the dataset and ensuring high-quality data selection and expert annotation by veterinary professionals.

Our gratitude extends to our NLP team, Aunchana Pimpisal and Punnawit Foithong, for their insightful discussions and feedback, as well as to the Axons Farm Business team for their efforts in translating this research into real-world applications for future improvements.

Finally, we appreciate Charoen Pokphand Foods (CPF) for initiating CPF Hack the Future, which brought us together—from the initial project pitch to this meaningful collaboration.

\bibliographystyle{unsrt}  
\bibliography{references}  

\end{document}